\documentclass{jetpl}
\twocolumn
\title{On the broken time translation symmetry in macroscopic systems:\\
 precessing states and off-diagonal long-range order}

\rtitle{On the broken time translation symmetry in macroscopic systems}

\sodtitle{On the broken time translation symmetry in macroscopic systems}

\author
{G.E. Volovik $^{+\#}$ \/\thanks{e-mail: volovik@boojum.hut.fi}
}

\rauthor{
G.E.Volovik
}

\sodauthor{
Volovik
}

\address{
$^{+}$ Low Temperature Laboratory, Aalto University, School of Science and
Technology, P.O. Box 15100, FI-00076 AALTO, Finland
\\
$^{\#}$ Landau Institute for Theoretical Physics RAS, Kosygina 2, 119334 Moscow, Russia
}

\abstract{ 
The broken symmetry state with off-diagonal long-range order (ODLRO), which is characterized by the vacuum expectation value of the operator of  creation of the conserved quantum number $Q$, has the time-dependent order parameter. However, the breaking of the time translation symmetry is observable only if the charge $Q$ is not strictly conserved and may decay. This dihotomy is resolved in systems with quasi-ODLRO.
These systems have two well separated relaxation times: the relaxation time $\tau_Q$ of the charge $Q$ and the energy relaxation time $\tau_E$. If $\tau_Q \gg \tau_E$, the perturbed system relaxes first to the state with the ODLRO, which persists for a long time and finally relaxes to the full equilibrium static state. In the limit $\tau_Q \rightarrow \infty$, but not in the strict limit case when the charge $Q$ is conserved, the intermediate ODLRO state can be considered as the ground state of the system at fixed $Q$ with the observable spontaneously broken time translation symmetry.
Examples of systems with quasi-ODLRO are provided by superfluid phase of liquid $^4$He, Bose-Einstein condensation of magnons (phase coherent spin precession) and precessing vortices.
}

\begin{document}

\maketitle

 \section{Introduction}

Let us consider systems characterized by  quasi-conserved macroscopic quantum number  $Q$. 
By quasi-conservation we mean  that the relaxation time $\tau_Q$ of the charge $Q$ is much larger than the relaxation time $\tau_E$ of energy. 
The quantum number $Q$ can be the particle number $N$ (number of baryons, atoms, etc.);  the number of quasiparticles (magnons, phonons, photons, kelvons. etc.); spin projection ${\cal S}_z$; angular momentum
projection ${\cal L}_z$; etc. 
In such systems the initial excited state first relatively rapidly relaxes to the state with minimal energy at fixed $Q$, and then relatively slowly relaxes to the state with equilibrium value $Q=Q_0$, where $dE/dQ=0$.  When $Q\neq Q_0$, the system may experience  oscillations with frequency $\omega=dE/dQ$ (if $Q$ is properly normalized). Such oscillations if exist  represent
the broken symmetry state characterized by the off-diagonal long-range order (ODLRO)
\begin{equation}
\left<\hat a^+ \right>  \propto e^{i\omega t + i\alpha}\,,
\label{SMF1}
\end{equation}
where $\hat a^+$ is the creation operator of the quantum number $Q$, with $\hat Q=\hat a^+\hat a$.

In the limit $\tau_Q\rightarrow \infty$, the state (\ref{SMF1}) can be considered as the state with spontaneously broken time translation symmetry as discussed in Ref. \cite{Wilczek2013}.
However, in the strict limit $\tau_Q= \infty$, when the charge $Q$ is conserved, the oscillations become unobservable, since the reference frame with respect to which they can be measured is lost. 
In other words,   when the charge $Q$ is strictly conserved the ground state oscillations are either absent 
\cite{Bruno2013a,Bruno2013b,Nozieres2013} or are not observable. We call the systems with $\tau_Q\gg \tau_E$, which experience the ODLRO in the intermediate time  $\tau_Q\gg t \gg \tau_E$, the systems with quasi-ODLRO.

Typical example of the system with quasi-ODLRO is the superfluid $^4$He, where $Q=N_4$ is the number of $^4$He atoms, and $\omega=\mu_4$ is the chemical potential. The life time of $^4$He atoms is finite due to proton decay, that is why in a full equilibrium the chemical potential is strictly zero, $\mu_{\rm full~equilibrium}=0$. The life-time of  the
 $^4$He atom  is astronomically large, $\tau_{Q}> 10^{34}$ years, that is why we are in the limit when  $N_4$ can be considered as conserved quantity where the $U(1)$ symmetry is obeyed and the chemical potential is well defined.  The ground state of the system with fixed $N_4$ is time dependent according to 
Eq.(\ref{SMF1}) with $\omega=\mu_4$. Observation of these oscillations is only possible if the $U(1)$ symmetry is explicitly violated and the proton decay is possibl. Thus the observation would serve as the experimental evidence for the non-conservation of the baryonic charge and the proton instability.

Below we consider some other examples of macroscopic systems with quasi-ODLRO, where $\tau_Q$ and $\tau_E$ are highly separated,  $\tau_Q \gg \tau_E$. These are magnon BEC in superfluid $^3$He-B and the precessing vortex system, also in superfluid $^3$He-B. Though in the present experimental situation the relaxation time $\tau_Q$ is not astronomically large (varying in $^3$He-B from minutes to hours), but in principle
the astronomical time can be reached at lower temperatures.
 
\section{Coherent spin precession and magnon BEC}

 %%%%%%%%%%%%%%%%%%%%%%%%%%%%%%%%%%%%%%%%%%%%%%%%%%%%%%%%%%%%%
\begin{figure}[htt]
 \includegraphics[width=0.5\textwidth]{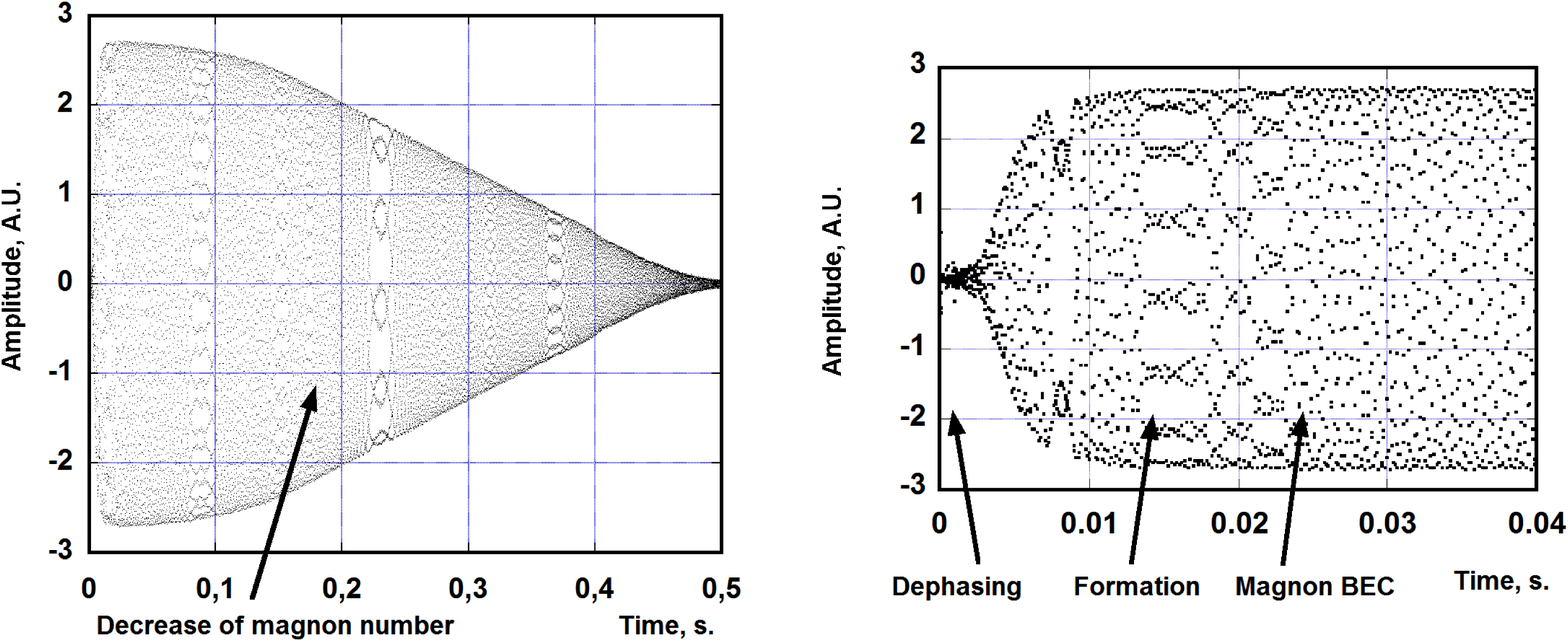}
 \caption{
 The stroboscopic record of  the free induction decay signal of coherent precession
 (from Ref. \cite{BunkovVolovik2013}). The same process discussed in the language of magnon BEC see in Fig. \ref{MagnonFig}.
{\it left}: During the first stage of about 0.002 s  the induction signal completely disappears due to dephasing. Then, during the energy relaxation time $\tau_E\sim 0.02$ s, the spin supercurrent redistributes the magnetization and creates the phase coherent precession, which is equivalent to the magnon BEC state. Due to small magnetic relaxation with $\tau_Q \sim 1$ s, the number of magnons slowly decreases but the precession remains coherent. Precession with essentially larger $\tau_Q$ emerges for magnon BEC confined in magneto-textural trap, see e.g. \cite{Autti2012}. 
  {\it right}: The magnon BEC signal for $t\ll \tau_Q$. Coherent spin precession looks as the ODLRO state with spontaneously broken symmetries: $U(1)$ symmetry (or equivalently the $SO(2)$ symmetry) and the time translation symmetry. 
 }
 \label{amplitude}
\end{figure}
%%%%%%%%%%%%%%%%%%%%%%%%%%%%%%%%%%%%%%%%%%%%%%%%%%%%%%%%%%%%%

In case when $Q={\cal S}_z$, the spin projection on the direction of magnetic field, the corresponding state with the off-diagonal long-range order represents the spontaneously emerging coherent spin precession \cite{BunkovVolovik2013} (see Fig. \ref{amplitude}).
  In a full equilibrium, the density of the spin projection  $S_z$ has equilibrium value, which  in magnetic field $H$ applied along 
  the axis $z$ is $S^z_{\rm full~equilibrium}=\chi H/\gamma$, where $\chi$ is spin susceptibility of $^3$He-B and $\gamma$ is the gyromagnetic ratio of the $^3$He atom. 
  In the limit, when losses of magnetization can be neglected, the spin projection ${\cal S}_z$ on magnetic field can be considered as conserved quantity, and the $SO(2)$ spin rotation symmetry is obeyed. For superfluid $^3$He-B one finds that  if it has the non-equilibrium value  of spin, ${\cal S}^z \neq V \chi H/\gamma$, it experiences spontaneous breaking of $SO(2)$ symmetry which is manifested by the phase coherent precession. This precessing state  has the ODLRO based on the operator of spin creation, $\left<S_+\right> =S\sin \beta   ~ e^{i\omega t+i\alpha}$, where $\beta$ is the tipping angle ($\cos\beta=S_z/S$); and the global frequency $\omega=dE/dS_z$  is coordinate independent even if the precession is spatially inhomogeneous due to textures and nonuniform magnetic field.

%%%%%%%%%%%%%%%%%%%%%%%%%%%%%%%%%%%%%%%%%%%%%%%%%%%%%%%%%%%%%
\begin{figure}[htt]
\includegraphics[width=\linewidth]{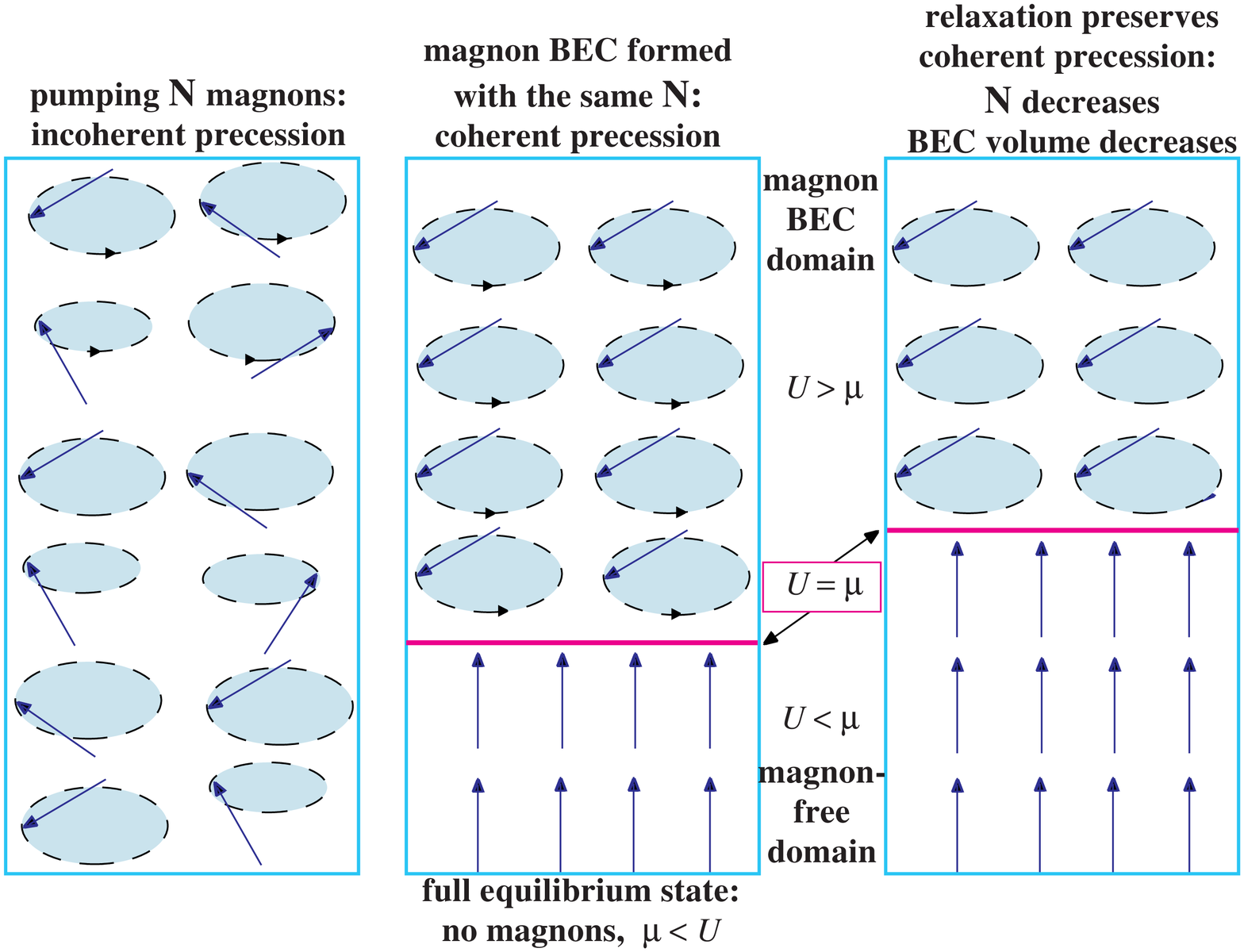}
\caption{Formation and relaxation of coherent spin precession in Fig. \ref{amplitude} in the language of the  Bose-Eistein condensation of magnons  (from Ref. \cite{BunkovVolovik2013}). {\it left}: Incoherent precession after applied pulsed NMR. {\it middle}: After short time $\tau_E$ of energy relaxation the ordered state is formed, which is represented by the BEC of magnons in the region where the chemical potential $\mu$ of magnons  exceeds the  external potential $U$. Magnon BEC is is manifested as precession with coherent phase and global frequency.  {\it right}: Relaxation of the quasi-conserved quantum number $Q={\cal S}_z$ does not destroy magnon BEC, but leads to the slow decrease of its volume.
}
 \label{MagnonFig}
\end{figure}
%%%%%%%%%%%%%%%%%%%%%%%%%%%%%%%%%%%%%%%%%%%%%%%%%%%%%%%%%%%%%

  In the limit  $\tau_Q \rightarrow \infty$ it is an example of the system with spontaneously broken time translation symmetry and off-diagonal long range order in the asymptotically free of dissipation regime, as discussed in Ref. \cite{Wilczek2013}.
  Note that the relaxation of the quantum number $Q$ does not destroy the phase coherent precession: it simply changes the volume of the region, where the precession takes place. 
The oscillations are observed through  the free induction decay signal generated by precession of magnetisation, i.e. they are observable because the interaction of spin with electromagnetic field explicitly violates the $SO(2)$ spin rotation symmetry.

  The relaxation time $\tau_Q$ in the infinite systems is mainly determined by spin-orbit interaction, which violates spin rotation symmetry and leads to the losses of spin. One can consider the model system with uniaxial anisotropy, in which the spin-orbit interaction does not violate the spin rotation symmetry along the $z$-axis. In this case the  ${\cal S}_z$ is good quantum number, and if the phase coherent precession is stable (the condition for that is $d^2E/dQ^2 >0$), the precessing state would correspond to the ground state at fixed $Q$. However, since in this model the charge $Q$ is strictly conserved, the connection with the orbital degrees of freedom is lost, and the time dependence is unobservable.

Actually the similar model in terms of two species has been considered in Ref. \cite{Wilczek2013}. 
The quantum number $Q$ there corresponds to the difference in populations, $Q=N_+-N_-$, while
the precession frequency is the difference in chemical potentials, $\omega=|\mu_+ - \mu_-|$. Here again if the charge $Q$ is fully conserved, the oscillations are not observable. For observability one must violate the $U(1)$ symmetry related to the conservation of charge $Q$, for example  by introducing the tunneling between the species. But in this case the oscillating state becomes non-equilibrium 
and will finally relax to the equilibrium state with $\mu_+=\mu_-$. The quasi-ODLRO state exists in this model if $\tau_Q \gg \tau_E$. The oscillations can be considered as the internal Josephson effect (though the internal Josephson effect discovered in $^3$He-A \cite{Webb1974} refers to the longitudinal oscillations of magnetization, i.e. oscillations of $S_z$ \cite{Leggett1975}).

The phase coherent precession has the parallel with the superfluid state of liquid $^4$He, which is explicitly seen if  one uses the language of magnons, see Fig. \ref{MagnonFig}. In this language, the quasi-conserved quantity $Q$ is the magnon number $N_M=({\cal S}-{\cal S}_z)/\hbar$, while the approximate $SO(2)$ spin symmetry is substituted by the approximate $U(1)$ symmetry in magnon system.  In the  full equilibrium, the chemical potential of magnons is zero, $\mu=0$, since their number is not conserved. The coherent precession in $^3$He-B corresponds to the  Bose-Einstein condensation of magnons (magnon BEC). The condition $\tau_Q\gg \tau_E$  in $^3$He-B means that the life time of magnon BEC is large compared to the time of thermalization, which leads to the formation of magnon BEC. In the limit of vanishing dissipation the magnon chemical potential  $\mu$ becomes well determined, being equal to the global frequency of precession, $\mu_M=\omega$, which in turn depends on the number $N_M$ of the pumped magnons. Relaxation slowly reduces the volume occupied by magnon BEC, but does not destroy it.

The related process in cosmology is discussed in Ref. \cite{Klinkhamer2012}.

\section{Vortex precession}

%%%%%%%%%%%%%%%%%%%%%%%%%%%%%%%%%%%%%%%%%%%%%%%%%%%%%%%%%%%%%
\begin{figure}[t]
\includegraphics[width=\linewidth]{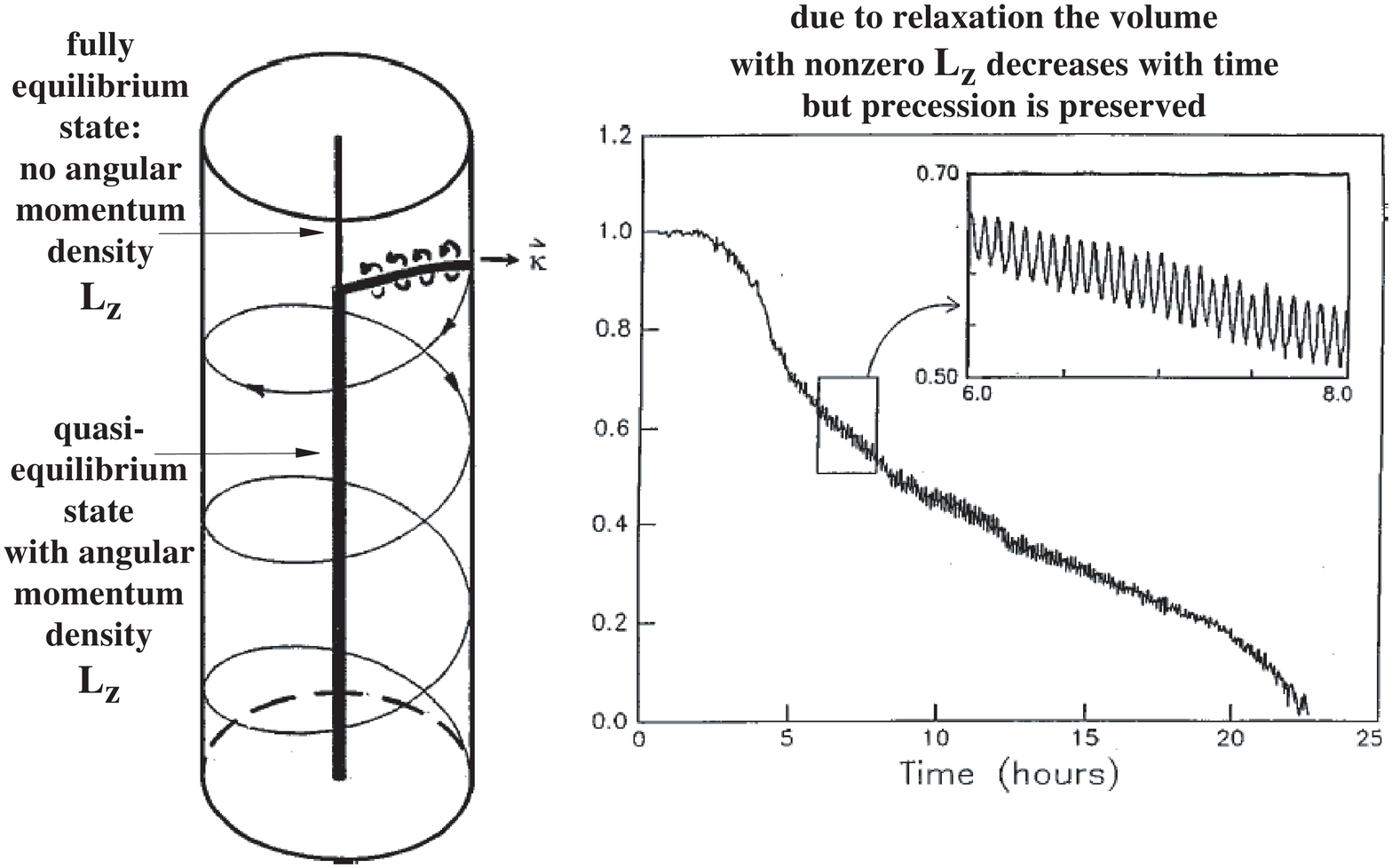}
\caption{Vortex precession and its relaxation (from Ref. \cite{Packard1998}).
{\it left}: Sketch of the partially trapped vortex. Circulation about the trapped part of the vortex 
gives rise to the orbital angular momentum, ${\cal L}_z$. The ${\cal L}_z$  is the quasi-conserved quantity, which is zero in the full equilibrium, when the vortex disappears. The nonzero value of  ${\cal L}_z$ in the quasi-equilibrium state, gives rise to the macroscopic precession with frequency $\omega=dE/d{\cal L}_z$. {\it right}: The further relaxation of ${\cal L}_z$ does not destroy precession, but decreases the length of the trapped part of the vortex (the vertical axis corresponds to the relative length of the trapped vortex). The process of relaxation is similar to that in Fig. \ref{amplitude} {\it left}.
}
 \label{VortexFig}
\end{figure}
%%%%%%%%%%%%%%%%%%%%%%%%%%%%%%%%%%%%%%%%%%%%%%%%%%%%%%%%%%%%%

Example of ODLRO is also provided by the precession of a vortex line  in $^3$He-B in Fig. \ref{VortexFig}, where hours long oscillations have been experimentally observed
\cite{Zieve1992,Packard1998}. The vortex is partially trapped by wire. Outside the trapped part of the vortex there is circulating flow of the liquid around the wire, which has the orbital angular momentum ${\cal L}_z=\hbar (\nu/2) n_3 V$, where $n_3$ is the density of $^3$He atoms; $\nu$ is the number of circulation quanta in the trapped vortex; $V=\pi R^2 l$ is the volume of the part of the container with trapped circulation; $l$ is the length of the trapped piece of the vortex.  ${\cal L}_z$ is not strictly conserved due to interaction with the static boundaries, but the interaction is highly reduced at low temperature, where  in the limit of small dissipation, 
the projection ${\cal L}_z$  can be considered as quasi-conserved quantum number.
The precessing state of a free part of the vortex can be represented in terms of the ODLRO based on the operator of creation of the orbital angular momentum
$\left<L_+\right> \propto  e^{i\omega t+i\alpha}$, where the precession frequency $\omega=dE/d{\cal L}_z$. 
The oscillations are observed due to explicit violation of the axial $SO(2)$ cylindrical symmetry: the wire, which traps the vortex,  is not exactly in the center of the cylindrical vessel. That is why the length of the  vortex is oscillating.

This coherent precession of a vortex can be described using the language of macroscopic ac Josephson effect
\cite{MisirpashaevVolovik1992} and also the language of the BEC: in this case the proper excitations are those which propagate along the vortex line -- Kelvin waves or kelvons.

The related process takes place in experiments with the propagating turbulent vortex front \cite{Hosio2011}, where the relaxation time of ${\cal L}_z$ is much longer than the energy relaxation time. As a result the front is also characterized by the precession of vortices. In the limit of low temperature the effect of the decoupling from the environment is observed, when the vortex system chooses its own angular velocity of rotation, which is independent of the angular velocity of container.

\section{Discussion}

 In conclusion, the broken time translation symmetry may emerge in the limit of large life time $\tau_Q$, but is absent at $\tau_Q =\infty$ when the charge $Q$ is strictly conserved, since the breaking of the time translation symmetry could be observable only through the decay of the charge $Q$. The compromise is reached in the systems with quasi-ODLRO.
We considered examples of such systems -- superfluid $^4$He, magnon BEC and precessing vortices. These systems have the quasi-conserved quantum number or charge $Q$, whose relaxation time  $\tau_Q$ is finite, but  is much larger than the energy relaxation time $\tau_E$.  In the time scale $\tau_Q \gg t \gg \tau_E$, these systems experience the symmetry breaking  characterized by the off-diagonal long-range order, with the nonzero vacuum expectation value of the operator of  creation of the quantum number $Q$.  Within this time interval the  ODLRO state represents the broken time translation symmetry in the ground state at fixed $Q$, or at fixed chemical potential $\mu$. 

It is the ODLRO, which distinguishes the considered states from the general periodic dynamical states, which can be also asymptotically free of dissipation. The class of the quasi-ODLRO states do not include for example the amplitude (Higgs) modes, which emerge after quench in superfluids and superconductors (see \cite{Gurarie2013,Matsunaga2013} and references therein) and oscillations emerging in cosmology (in inflationary models \cite{Starobinsky1980}; in the models of the vacuum energy decay  \cite{KV2009}; in cyclic Universes \cite{BarsSteinhardtTurok2013}; etc.). 

\section*{\hspace*{-4.5mm}ACKNOWLEDGMENTS}
I acknowledge financial
support  by the EU 7th Framework Programme
(FP7/2007-2013, grant $\#$228464 Microkelvin) and by the Academy of
Finland through its LTQ CoE grant (project $\#$250280).

 \end{document}